\documentclass[reprint]{revtex4-1}

\usepackage{graphics}
\usepackage{graphicx}
\usepackage{amssymb}
\usepackage{amsthm}
\usepackage{amsmath}
\usepackage{multirow}
\usepackage{soul}
\usepackage{xcolor}
\usepackage{verbatim}
\usepackage{color}

\begin{document}

\title{Shear and crystallization in deformable granular packings: why don't auxetics order?}

\author{Joel T. Clemmer}
\email{Correspondence to: jtclemm@sandia.gov}
\author{Nicholas W. Hackney}
\author{Gary S. Grest}
\affiliation{Sandia National Laboratories, Albuquerque, New Mexico 87185, USA}
\date{\today}

\begin{abstract}
Shear of three-dimensional, highly compressed granular packings is simulated using a bonded particle approach that explicitly resolves elastic deformation.
Varying Poisson's ratio $\nu$ produces significant changes in  rheology, packing structure, and grain morphology.
During flow, conventional systems ($\nu > 0$) readily crystallize while auxetics ($\nu < 0$) resist ordering.
This duality reflects the fact that conventional grains develop polyhedral-like facets but conserve volume while auxetics behave oppositely, demonstrating an unexpected interaction between elasticity, geometry, and crystallization.
\end{abstract}

\maketitle



Granular materials are often studied in the rigid limit where pressures are small relative to elastic moduli such that deformation of individual grains is negligible.
However, there is growing interest in the physics of granular packings at higher pressures due to varied applications from fragmentation in fault gouges \cite{Marone1989, Clemmer2022}, pharmaceutical tableting \cite{Jain1999}, and the crushing of wild blueberries in packaging and processing \cite{Forney2006}.
There is also the possibility that mechanisms like deformation can introduce new physical phenomena \cite{Manning2023}.
Here we focus on elastically deformable soft/squishy granular matter \cite{Bares2022} that can include silicone \cite{Bares2023} or hydrogel \cite{Dijksman2022, Tapia2024, Fan2025} beads, granular metamaterials \cite{Shim2012, Haver2024}, curved colloidal liquid crystals \cite{Hackney2026}, and biological cells \cite{Bi2014}.

In the rigid limit, the structure \cite{Donev2004, Monti2022}, mechanical properties \cite{Athanassiadis2014}, and rheology \cite{Trulsson2018, Salerno2018} of granular materials are all intrinsically tied to the geometry (size and shape) of the individual grains. 
Three-dimensional monodisperse spheres cannot pack more densely than a face centered cubic (FCC) crystal.
However, spheres can get trapped in less dense disordered glassy states due to a mismatch in drivers for local and global structure \cite{Nelson1983, Hoy2024}. 
This tendency can then be manipulated by perturbing geometry.
For example, polydispersity can further inhibit crystallization \cite{Zaccarelli2009}, while polyhedral grains can promote crystallization or even change the crystal structure \cite{Damasceno2012, Newman2019}.
When incorporating elastic deformation, grain geometries can then evolve leading to largely unknown effects on the macroscopic response.

\begin{figure*}
	\includegraphics[width=0.95\textwidth]{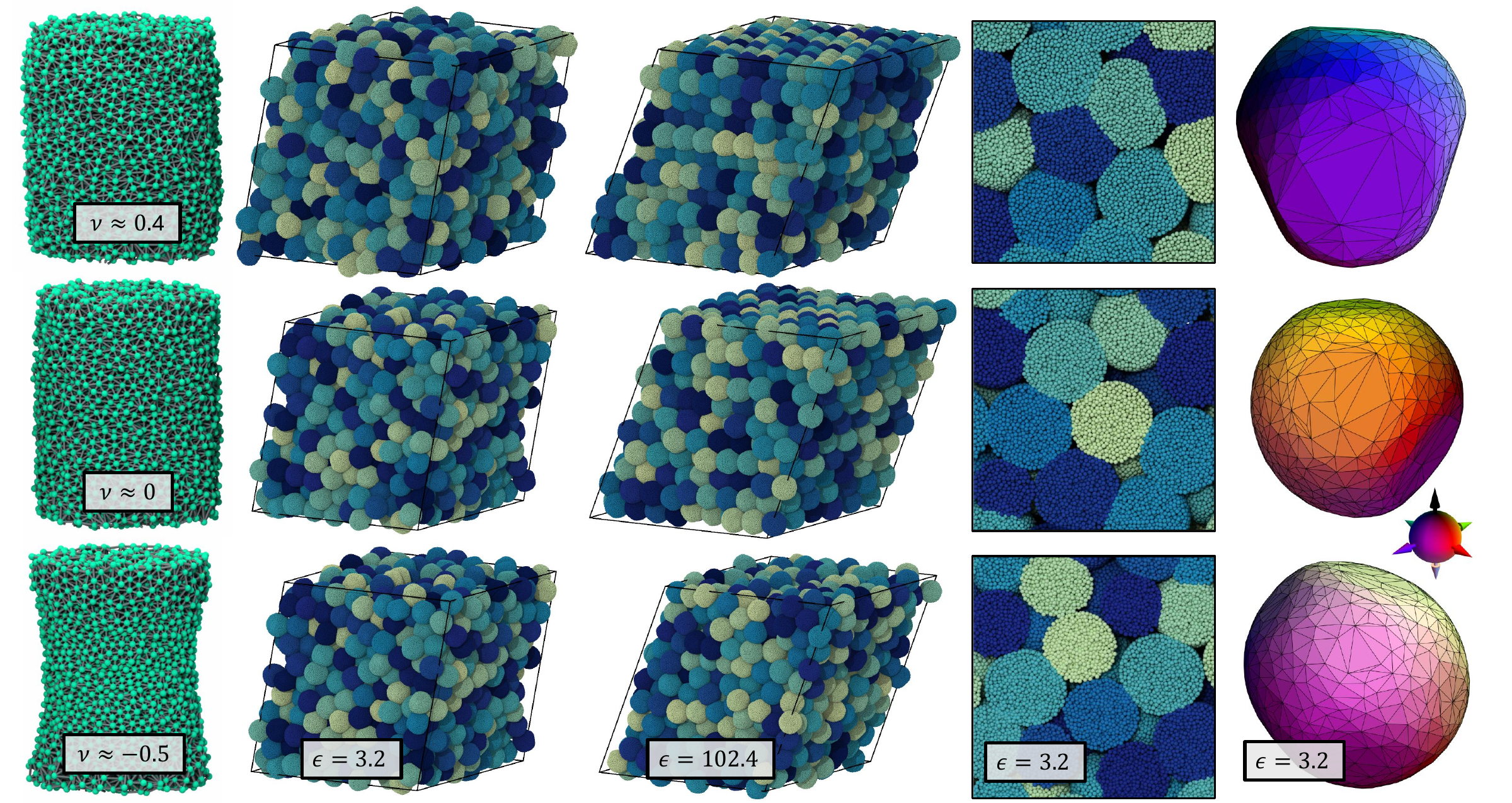}
	\caption{Rendered simulations with Poisson's ratios of $\nu = 0.4$ (top), $0$ (middle), and -0.5 (bottom). Column 1: example cylinders undergo uniaxial compression to demonstrate the Poisson's effect. Columns 2 and 3: sheared granular systems at strains $\epsilon = 3.2$ and 102.4, respectively. Column 4: zoomed in cross sections at $\epsilon = 3.2$. In columns 2-4, all nodes in a grain share a color. Column 5: convex hulls of individual grains are rendered as a triangle mesh. Color denotes the orientation of each triangle's normal vector according to the included guide.}
	\label{fig:visuals}
\end{figure*}

In isotropic linear elasticity, mechanics are governed by the Young's modulus $E$ and Poisson's ratio $\nu$ which roughly determine the magnitude and nature of deformations, respectively.
In conventional materials ($\nu > 0$), compressing an object along one axis causes the object to bulge outward laterally in an effort to minimize volumetric changes.
At its max $\nu = 1/2$, a material is incompressible.
The extent of this bulging decreases as $\nu$ is reduced until the object no longer laterally distorts at $\nu = 0$, akin to a cork \cite{Gibson1981}.
For $\nu < 0$, materials are classified as auxetic and the lateral dimensions shrinks inwards upon axial compression (demonstrated in Fig.~\ref{fig:visuals}).
An auxetic object can thus be thought of as trying to minimize changes in shape at the cost of changing volume. 
This Poisson's effect provides a simple but curious tool for controlling geometric changes in soft/squishy materials. 

To investigate such scenarios of varying $\nu$, we use a bonded particle model (BPM) to simulate deformable granular material at high pressure under shear. 
At small strains, disparities emerge in grain geometries as flow is marked by rapid fluctuations and broad distributions in the shape of individual grains in the conventional limit ($\nu > 0$) or volume in the auxetic limit ($\nu < 0$).
At large strain, grains with conventional elasticity readily crystallize as grains distort into polyhedral-like shapes that promote this transition.
Such ordering is comparatively rare in low-pressure systems.
In contrast, auxetic grains tend to remain disordered as they naturally become more dispersed in volume.
This highlights a dichotomy between the contrasting effects of evolving shape vs. size of grains in conventional and auxetic materials that connect elastic deformation, rheology, and ordering. 


While defining isotropic linear elasticity requires two parameters, say $E$ and $\nu$, in traditional discrete element method (DEM) simulations these parameters only set the stiffness of contact forces as grains are modeled as rigid objects.
Alternative methods, such as the BPM used here, explicitly represent elastic deformation of grains such that the values of $E$ and $\nu$ are independently important \cite{Cardenas-Barrantes2022, Silling2023, Nezamabadi2024}.
We simulate 1024 initially monodisperse and spherical grains in a cubic, periodic box using the BPM package in LAMMPS \cite{Thompson2021, Clemmer2024}, which specializes in force models that depend on a reference configuration.
Each node has a diameter of $d$ and mass of $m$ (the units of length and mass) with numerically updated positions and velocities similar to traditional molecular dynamics.
Each grain is resolved using $5137$ nodes (computational particles) in a disordered configuration with an undeformed radius of $R \approx 10 d$.

Within each grain, all neighboring nodes less than a distance of $1.5d$ are connected by a bond.
The magnitude of the force between bonded nodes $i$ and $j$ is given by
\begin{equation*}
    k \left(r_0 - r \right) + \alpha \left( \left[ \frac{V_i + V_j}{V_{0,i} + V_{0,j}} \right]^{\frac{1}{3}} - 
                  \frac{r}{r_0} \right) - \sqrt{k m} \hat{r} \cdot \vec{v}
    \label{eq:bond}
\end{equation*}
where $r$ and $r_0$ are the current and initial distances between the two nodes, $\hat{r}$ is the normalized displacement, $k$ is a stiffness, $\alpha$ is a fitting term (described latter), $V_i$ and $V_{0,i}$ are the current and initial volumes of node $i$, and $\vec{v}$ is the velocity difference.
We approximate $V_i \propto \sum r^3$, summing across all bonded neighbors $j$, akin to a many-body embedded atom potential. 
The third term is a damping force to remove excess kinetic energy.
Unbonded nodes within contact (a distance $<d$) exchange a similar Hookean and damping force as well as an anharmonic cubic repulsion with a prefactor of $50 k/d^{2}$ to ensure grains do not overlap a distance greater than $d/2$.

The two free parameters $k$ and $\alpha$ are used to specify Young's modulus $E$ and Poisson's ratio $\nu$. 
Here, we choose $k$ to be the third fundamental unit which defines a unit of time $\tau = \sqrt{m/k}$ (and a timestep of $0.1\tau$) while the many-body $\alpha$ sets $\nu$
\footnote{Instead of fixing $k$, one could alternatively fix Young's modulus $E$ which would amount to a unit change.}.
Under isotropic compression, all bonds are similarly compressed leading to minimal differences between per-bond and volumetric strains in the many-body term such that $\alpha$'s value is largely irrelevant.
However, under shear, bonds are both stretched and extended such that $\alpha$ is relevant and thus controls the resistance to shear.
At $\alpha = 0$, the force reduces to a pairwise interaction which constrains $\nu = 1/4$ \cite{Greaves2013}.
Positive $\alpha$ increases shear resistance (decreasing $\nu$) and negative $\alpha$ does the opposite.
This control over $\nu$ is demonstrated by creating cylindrical geometries and compressing their top and bottom sections to induce the lateral deformations seen in Fig. \ref{fig:visuals}.
We consider values of $\alpha =$ -0.8, 0, 2, 5, and 12 corresponding to $\{\nu, E d/k\}$ of  \{0.40, 0.58\}, \{0.25, 1.84\}, \{-0.01, 4.05\}, \{-0.25, 6.04\}, and \{-0.52, 8.30\} using the calibration defined in Ref.~\cite{Clemmer2024}. 
For convenience, we round values of $\nu$ to -0.5, -0.25, 0.0, 0.25, and 0.4.


To create packings for shear, three system realizations are generated below jamming.
Their simulation cells are then isotropically compressed to near full density at high pressure $P$ while affinely remapping nodal positions at each value of $\nu$. 
Stress tensors are calculated by summing both the virial and kinetic energy contributions across all nodes and normalizing by the box volume.
Packing fractions $\phi$ are measured with Monte Carlo sampling.
Collisions are checked with both nodal spheres and the convex hull of grains, accounting for both overlaps and inter-node gaps. 
To facilitate comparisons across $\nu$, pressures $P$ are normalized by the effective Hertz modulus which is the stiffness of contact forces at small overlaps:  $\tilde{P} \equiv P(2 - 2\nu^2)/E$.
This effectively collapses compaction curves of $\phi$ vs. $\tilde{P}$ at low pressures. 
Systems jam at $\phi_c \approx 0.6$ above which $\tilde{P}$ grows according to the expected power law of $(\phi - \phi_c)^{3/2}$ \cite{OHern2003} until $\Delta \phi > (\phi - \phi_c) \approx 0.1$ or $\tilde{P} \approx 10^{-2}$ above which the growth in $\tilde{P}$ accelerates and elastic deformation in grains becomes significant \cite{Clemmer2024}.

\begin{figure}
	\includegraphics[width=0.85\columnwidth]{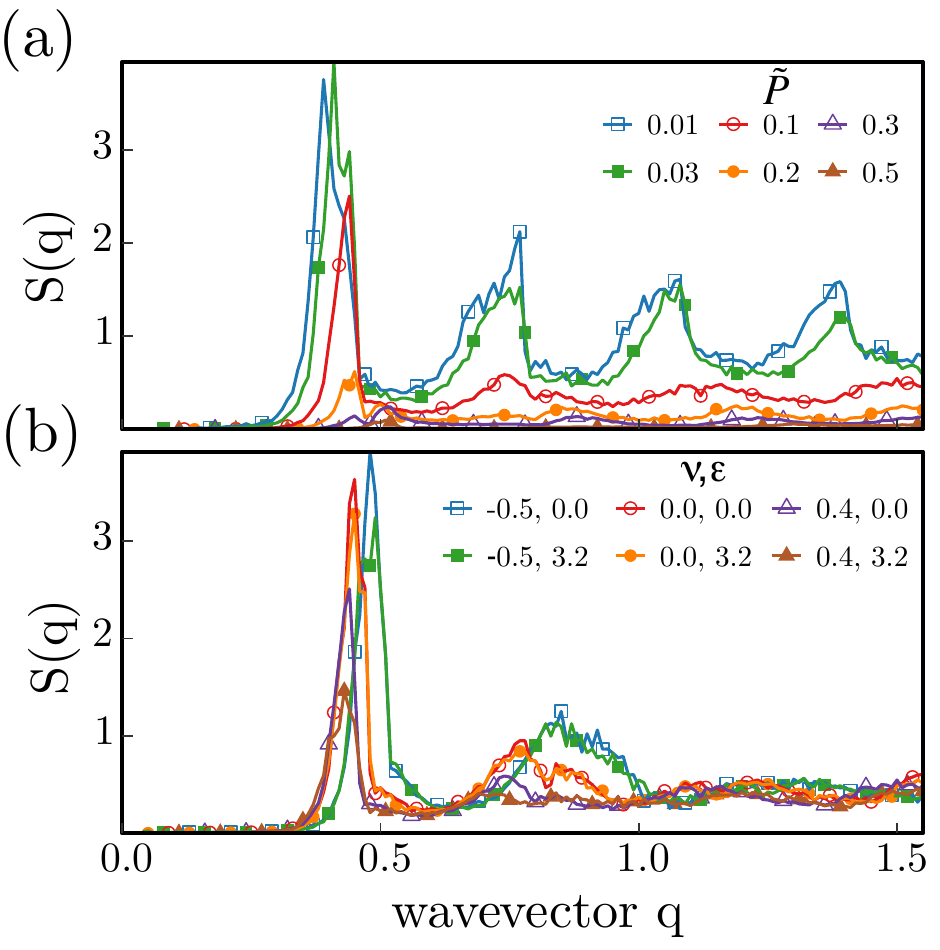}
	\caption{(a) Structure factor $S(q)$ as a function of wavevector $q$ (units $1/d$) during initial compaction at the indicated pressures $\tilde{P}$ at $\nu = 0.4$. (b) Structure factor at $\tilde{P} = 0.1$ at the indicated Poisson ratio $\nu$ and strains $\epsilon$.}
	\label{fig:packing_structure}
\end{figure}

Structural changes above jamming are quantified in terms of the spherically-averaged structure factor $S(q)$ of the packing.
$S(q)$ is calculated by first computing the form factor $P_{F,i}(q)$ for each individual grain (Eq. S1) and averaging to obtain $P_F(q) \equiv \langle P_{F,i}(q)\rangle$ (Fig. S1).
For very low $\tilde{P}$, $P_F(q)$ is well-described by the form factor for a perfect sphere \cite{Gommes2021} of radius $R=10d$. 
As $\tilde{P}$ increases, the distinctive structure of $P_F(q)$ for a sphere is suppressed as the individual grains deform.
The static structure factor $S(q) \equiv I(q) / P_F(q)$, where $I(q)$ is the total scattering intensity of all the nodes in the system (Eq. S2). $S(q)$ highlights the transition from a typical packing of monodisperse spheres at relatively low $\bar{P} = 10^{-2}$ ($\phi \approx 0.68$) to an almost fully dense solid ($\phi \approx 0.99$) with minimal structure at the highest $\bar{P} = 0.5$ [Fig. \ref{fig:packing_structure}(a)].
For the remainder of this letter, we focus on $\tilde{P} = 0.1$ as systems exhibit significant deviations from packings of rigid spheres while retaining recognizable discrete structure. 

After compaction, the simulation cell is triclinically tilted to apply simple shear at a nearly quasistatic strain rate of $1.6 \times 10^{-5} \tau^{-1}$ (Fig. S3).
During shear, pressure is maintained by isotropically dilating or contracting the box using a linear controller with a gain of $10^{-5} k/d$, akin to deformation protocols performed in studies of low-pressure granular rheology \cite{Srivastava2020, Clemmer2021}. 

First we consider results before shear and up to relatively small strains $\epsilon$ of less than $5.0$ units.
In both cases, the packing fraction $\phi$ has a slight dependence on $\nu$ as $\phi$ generally drops from from $\approx 0.87$ to $\approx 0.85$ as $\nu$ decreases from 0.4 to -0.5.
Secondly, there are significant differences in structure between conventional ($\nu > 0$) and auxetic ($\nu < 0$) grains, quantified in Fig. \ref{fig:packing_structure}(b).
Visually, conventional grains exhibit significant flattening at interfaces creating polyhedral-like facets while auxetic grains remain quite spherical in Fig. \ref{fig:visuals}. 
This preservation of shape leads to significantly more structure in $S(q)$ being preserved in the auxetic limit.
The relative difference in $S(q)$ across $\nu$ persists and is even accentuated during shear as even more packing structure is lost at $\nu = 0.4$.
Here, $S(q)$ is spherically averaged as a first approximation to compare to the initial state, as all systems are disordered at $\epsilon = 3.2$.

\begin{figure}
	\includegraphics[width=\columnwidth]{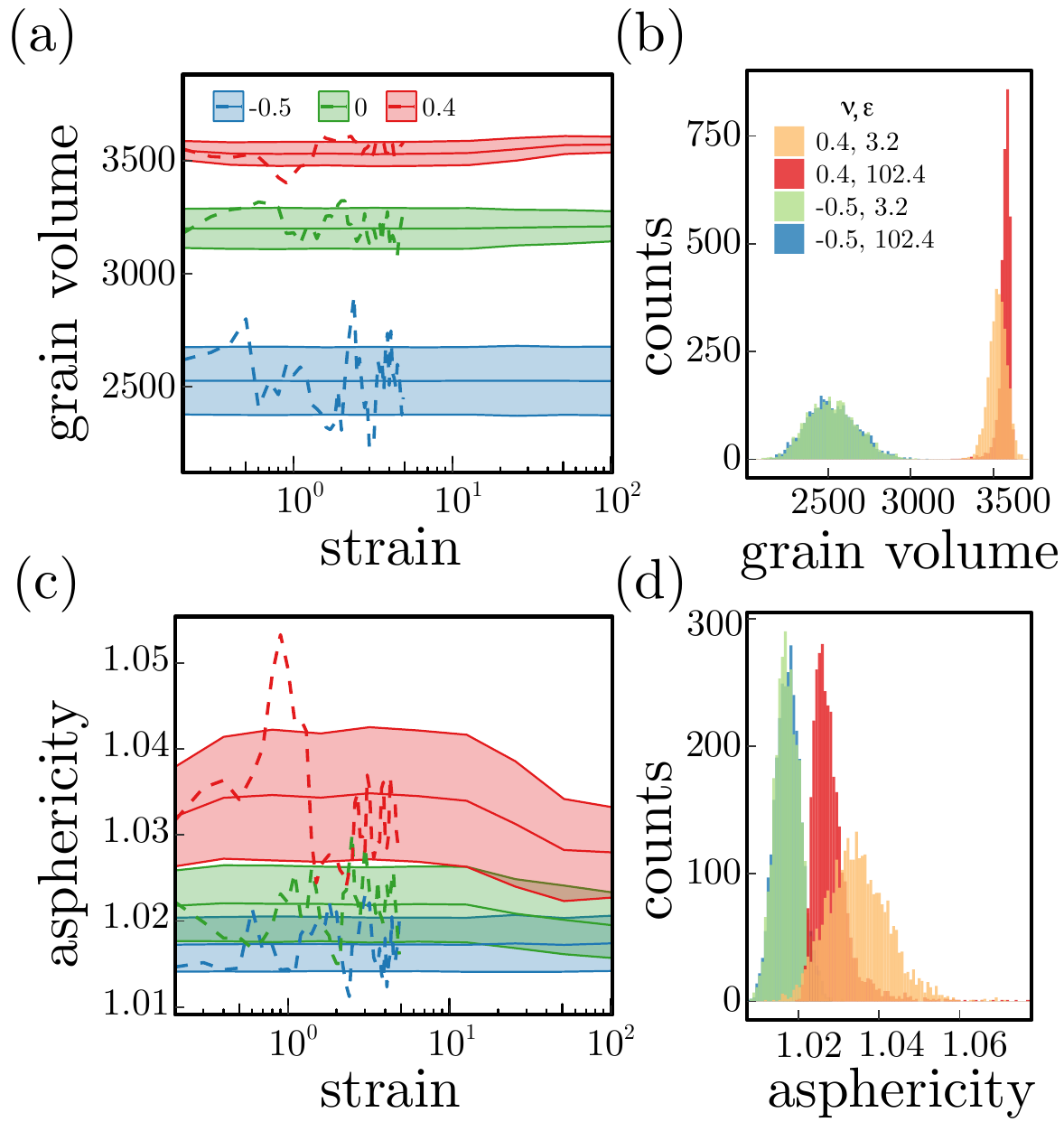}
	\caption{(a,c) Evolution in the average granular volume (units $d^3$) and asphericity with strain for the indicated $\nu$. Ribbons demark one standard deviation and dashed lines are traces of a single grain. (b, d) Distributions of the number of grains of a given volume and asphericity.}
	\label{fig:grain_geometry}
\end{figure}

\begin{figure*}
	\includegraphics[width=\textwidth]{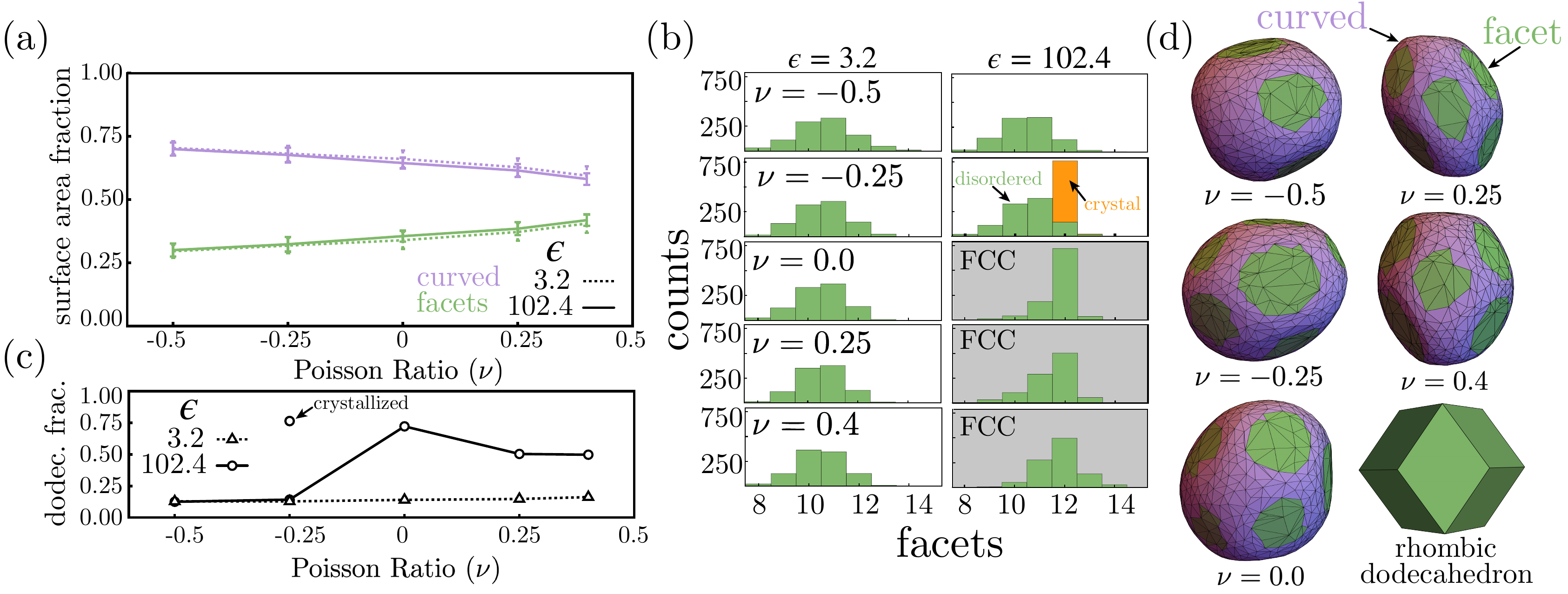}
	\caption{(a) Average fraction of surface area in faceted (green) or curved (purple) domains across $\nu$ at the indicated $\epsilon$. (b) Distributions of the number of facets where shaded boxes indicate crystallized states. Both histograms in one row correspond to the same value of $\nu$, as labeled in the left-most column. In the specific case of $\nu=-0.25$ and $\epsilon=102.4$, results for two simulations provided: one corresponding to a crystallized configuration (orange) and one that is disordered (green). (c) The fraction of dodecahedron-like grains with 12 facets versus $\nu$ at small (dashed) and large (solid) strains. Data for $\nu = -0.25$ for systems that remained disordered vs. crystallized are separately labeled. (d) Example curved and faceted regions of grains across $\nu$ are shown along with a rhombic dodecahedron for comparison.}
	\label{fig:facets}
\end{figure*}

In addition to structure, significant changes in rheology are identified with varying Poisson ratio.
At $\nu = 0.4$, the average stress ratio is found to be $\mu = 0.17$, where $\mu$ is defined as the shear stress in the direction of tilt normalized by $P$ (Fig. S2-3). 
This value of $\mu$, a common measure of granular rheology that can be interpreted as an internal friction coefficient \cite{Jop2006}, is notably small in comparison to values of $\mu$ measured in the rigid limit \cite{Srivastava2020, Clemmer2021}.
This may reflect the fact that deformation enables new mechanisms for local rearrangement, such as a flattening of contacts to allow grains to slide past one another. 
Even more remarkable, however, is that $\mu$ further decreases to $0.1$ for $\nu = -0.5$. 
This is on par with values of $\mu$ for frictionless spheres in the rigid limit.
Since auxetic grains have relatively less facetting and thus smaller contact areas, they may additionally reduce frictional forces and thus $\mu$.
\citet{Haver2024} observed a similar $\nu$-dependence of $\mu$ in experimental oscillatory flows of two dimensional disk packings, as auxetic grains can dynamically reduce their volume to slip past small gaps during flow. 
\citet{Haver2024} also reported more frequent fluctuations in auxetics, in agreement with our simulations (Fig. S2).

The convex hull of each grain (as seen in Fig.~\ref{fig:visuals}) is used to calculate both its volume $V$ and surface area $A$.
The average volume of a grain is larger in conventional materials (less compaction), however, variations in volumes are larger in auxetic systems [Fig.~\ref{fig:grain_geometry}(a)].
Furthermore, tracing individual auxetic grains reveals there are also rapid temporal fluctuations in their volume.

To measure distortions in shape, one can  define a measure of asphericity $A^{3/2} / (6 \sqrt{\pi} V)$ which equals unity for a sphere and increases as the shape changes \cite{Boromand2019, Cardenas-Barrantes2022}. 
The dependence of asphericity on $\nu$ behaves oppositely to the volume, as conventional materials have larger distortion and fluctuations [Fig.~\ref{fig:grain_geometry}(c)]. 
Furthermore, as strains increase above $\epsilon = 10.0$, the average asphericity exhibits a peculiar drop at $\nu = 0.4$ and $0.0$ which is not present at $\nu = -0.5$.
As shown in Fig.~\ref{fig:grain_geometry}(b,c), significant differences in distributions of both the volume and asphericity are also present between small and large strains at $\nu = 0.4$ while no detectable change is seen at $\nu = -0.5$. 
From the $\epsilon = 102.4$ images in Fig.~\ref{fig:visuals}, one can easily see this transition is due to crystallization.
By this strain, we find crystallization never occurs at $\nu = -0.5$, occurs in 1 out of 3 samples for $\nu = -0.25$, and always occurs at $\nu \ge 0$.
As this crystallization triggers a large drop in the average asphericity of a grain but not the volume, it suggests the transition must specifically release shear stress within grains.
However, as auxetic materials resist shear deformation, this transition is much less critical.

To further characterize how shape distortion could drive crystallization, distinct facets of the grains are identified.
Using the convex hull, each triangle $i$ on the mesh compares its normal vector $\hat{n}_i$ to all abutting triangles $j$.
Triangles with aligned normal vectors ($\hat{n}_i \cdot \hat{n}_j > 0.99$) are grouped into the same facet. 
Comparisons repeat until the entire facet is identified. 
If a triangle misaligns with all neighbors, it is classified as belonging to a curved domain.
Example partitions are seen in Fig.~\ref{fig:facets}(d).

At both small ($\epsilon = 3.2$) and large strains ($\epsilon = 102.4$), the amount of surface area in facets consistently grows with $\nu$ [Fig.~\ref{fig:facets}(a)], reflecting observations of asphericity.
More telling, however, are distributions of the number of distinct facets in grains in disordered states for $\nu < 0$ and crystallized states for $\nu \ge 0$ [Fig.~\ref{fig:facets}(b)].
In disordered states, distributions are broader and have a peak between 10 and 11. 
After crystallization, however, there is a much stronger peak at 12 facets, the expected number of neighbors in an FCC crystalline lattice.
This is more clearly seen in the fraction of grains with 12 facets, as in a dodecahedron, which has no dependence on $\nu$ at small strains but grows significantly as a function of $\nu$ at large strains [Fig.~\ref{fig:facets}(c)]. 
From visual inspection, the shape of grains at large $\nu$ even resembles a rhombic dodecahedron, the shape of a voronoi cell in an FCC lattice.
Studies by \citet{Newman2019} on the crystallization pathways of different rigid polyhedra found that the barriers to crystallization shrink as an object reduces its spherical roundness, approaching a rhombic dodecahedron.
Therefore, we hypothesize that the deformation of conventional grains produces a less round geometry that more closely resembles rhombic dodecahedron, reducing barriers to crystallization.
As auxetic grains resist changes in shape and even develop significant polydispersity in volume, barriers remain high.


Shear-induced crystallization at high pressures has been studied before in simulations of wet suspensions of rigid spheres at high pressures by \citet{Khabaz2017} implying higher pressures and larger overlaps fundamentally increase the tendency to order. 
However, this work is the first time high pressure shear is accurately modeled by explicitly representing elastic deformation. 
While the preferred state for monodisperse spheres is expected to be ordered, finding such a state is generally a much slower process.
For instance, we ran analogous DEM simulations at a low pressure ($\tilde{P} = 10^{-5}$) to strains of 1000 and crystallization was never identified. 
Furthermore, the dependence of this transition on $\nu$ (which cannot be represented in DEM) provides insight into its nature and highlights the physical consequences of auxetic elasticity.

These results demonstrate a unique convergence of different physical effects/properties that emerge at high pressures through elastic deformation.
Despite the rarity of auxetic materials, such behaviors are still accessible by designing metamaterial grains that mimic auxetic elasticity \cite{Shim2012, Haver2024}.
Furthermore, these results provoke questions of whether one could design structures that further enhance or suppress the observed crystallization.
For instance, could one design a grain that converts its structure between a sphere and a rhombic dodecahadron with pressure? 
If so, could one design granular matter that readily orders with small changes in confinement?
What might happen if grains initially had polyhedral shapes or if deformation included irreversible mechanisms like plasticity or fracture \cite{Clemmer2021, Clemmer2025}?
This is a clear example of how deformation leads to a wide realm of possibilities in soft matter \cite{Manning2023}. 

\vspace{\baselineskip}
\textit{Acknowledgments} --
This work is funded by the Advanced Simulation and Computing program.
This work was performed, in part, at the Center for Integrated Nanotechnologies, an Office of Science User Facility operated for the U.S. Department of Energy (DOE) Office of Science. 
Sandia National Laboratories is a multi-mission laboratory managed and operated by National Technology \& Engineering Solutions of Sandia, LLC (NTESS), a wholly owned subsidiary of Honeywell International Inc., for the U.S. Department of Energy’s National Nuclear Security Administration (DOE/NNSA) under contract DE-NA0003525. This written work is authored by an employee of NTESS. The employee, not NTESS, owns the right, title and interest in and to the written work and is responsible for its contents. Any subjective views or opinions that might be expressed in the written work do not necessarily represent the views of the U.S. Government. The publisher acknowledges that the U.S. Government retains a non-exclusive, paid-up, irrevocable, world-wide license to publish or reproduce the published form of this written work or allow others to do so, for U.S. Government purposes. The DOE will provide public access to results of federally sponsored research in accordance with the DOE Public Access Plan.


\bibliographystyle{apsrev4-1}
\bibliography{bibliography.bib}

\newpage
\clearpage
\newpage

\onecolumngrid
\begin{center}
    \large\bfseries Supplemental information\\Shear and crystallization in deformable granular packings: why don't auxetics order?
\end{center}
\twocolumngrid

\setcounter{page}{1}
\setcounter{section}{0}
\setcounter{figure}{0}
\renewcommand{\thefigure}{S\arabic{figure}}
\renewcommand{\thepage}{S\arabic{page}}


\section{Form Factor}

\begin{figure}
	\includegraphics[width=0.9\columnwidth]{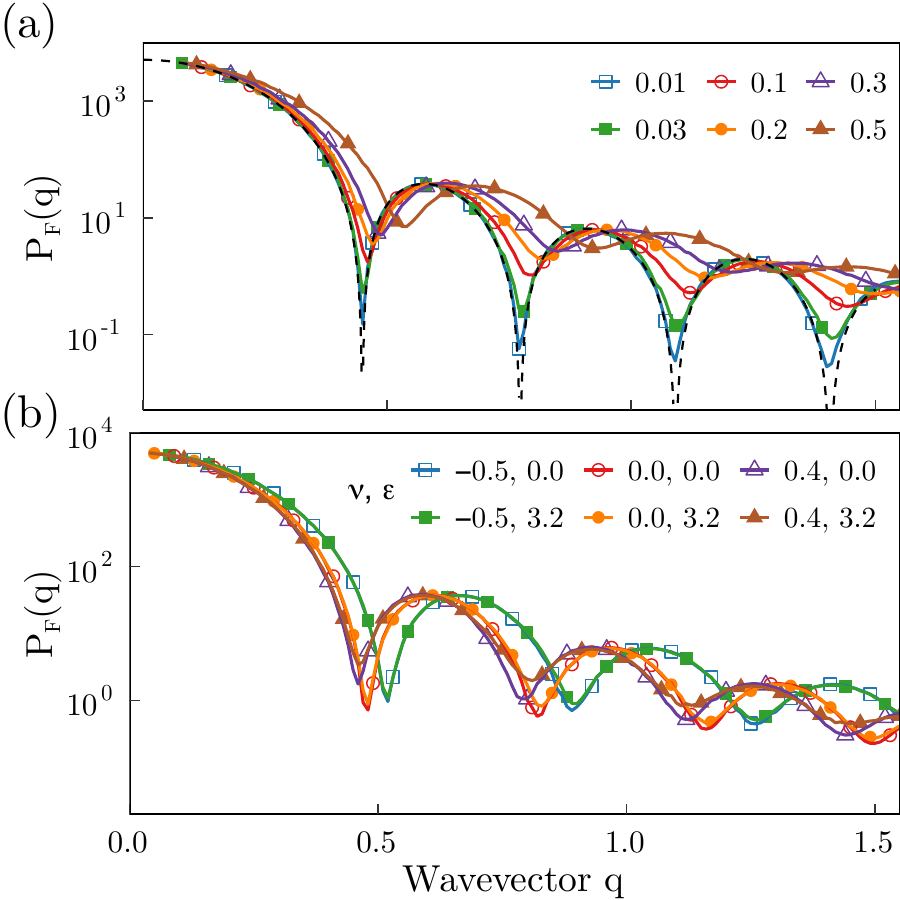}
	\caption{(a) Average form factor $P_F(q)$ for a grain at the initial compaction at the indicated pressures $\tilde{P}$ for $\nu = 0.4$. (b) $P_F(q)$ at the indicated Poisson ratio $\nu$ and strains $\epsilon$ for $\tilde{P} = 0.1$. The dashed, black line in (a) is the form factor for a perfect sphere of radius $10d$.}
	\label{fig:form_structure}
\end{figure}

The form factor $P_{F,k}(q)$ for grain $k$ is given by 
\begin{equation}
    P_{F,k}(q) = \frac{1}{N_k} \sum_{i,j=1}^{N_k} e^{i \vec{q} \cdot (\vec{r}_i - \vec{r}_j)}
\end{equation}
where $q$ is the momentum transfer vector, $\vec{r}_i$ is the position vector of node $i$, and $N_k$ is the number of nodes in the grain.  
Results for the average form factor $P_F(q) = \langle P_{F,k}(q) \rangle$ are shown in Fig.~\ref{fig:form_structure} for the initial compaction at the indicated normalized pressures $\tilde{P} \equiv P/E_\mathrm{eff}$ for a Poisson's ratio of $\nu = 0.4$ in panel (a) where $E_\mathrm{eff}$ is the effective Hertz modulus.
In panel (b), form factors are shown for the indicated $\nu$ and strains $\epsilon$ for $\tilde{P} = 0.1$.  
The dashed, black line in the first panel shows the results for a perfect sphere \cite{Gommes2021} of radius $R = 10d$, which is in good agreement with the measured $P(q)$ for low pressures.

The total scattering intensity $I(q)$
\begin{equation}
    I(q) = \frac{1}{\mathcal{N}} \sum_{i, j=1}^{\mathcal{N}} e^{i \vec{q} \cdot (\vec{r}_i - \vec{r}_j)}
\end{equation}
where $\mathcal{N}$ is the total number of nodes in the sample.  
Due to the periodic boundary conditions, $q$ values for calculating $I(q)$ are limited to $q =  \frac{2\pi}{L}  (n_x,n_y,n_z)$ where $L$ is the linear system size and $n_x$, $n_y$, and $n_z$  are integers. 
Since we are interested in the structure factor $S(q) = I(q)/P(q)$, the same $q$ values are used to determine $P(q)$. 
Details of calculating $I(q)$ for the triclinic simulation cell for $\epsilon=3.2$ are given in \citet{Monti2023}.

\section{Shear Response}

The response to shear stress is measured as the stress ratio $\mu = \sigma / P$ where $\sigma$ is the stress in the direction of the box tilt and $P$ is the confining pressure.
This stress ratio $\mu$ is a common metric in granular rheology that scales out the pressure-dependence in the rigid limit \cite{Jop2006}.
Here, all simulations are run at a target pressure $\tilde{P} = 0.1$.
Deviations in pressure due to the barostat are generally less than 0.1\% of the target.

\begin{figure}
	\includegraphics[width=0.9\columnwidth]{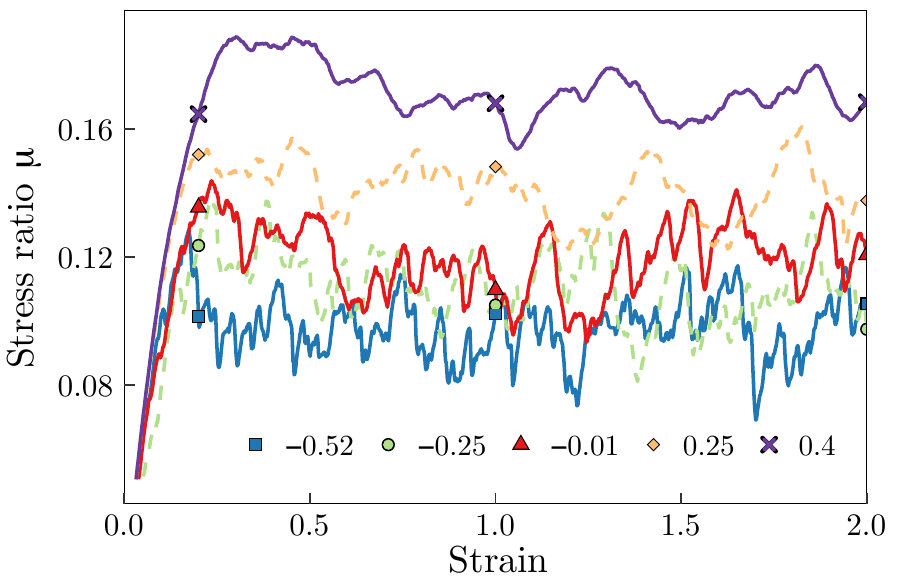}
	\caption{The stress ratio $\mu$ as a function of strain for one sample for the indicated values of the Poisson's ratio $\nu$ at $\tilde{P} = 0.1$ and $\dot{\epsilon} = 1.6 \times 10^{-5} \tau^{-1}$.}
	\label{fig:stress_strain}
\end{figure}

As a packing shears, it initially loads elastically before yielding and entering a plastic flow regime where $\mu$ fluctuates around an average value (Fig.~\ref{fig:stress_strain}).
This flow regime is not always an actual steady state as some systems eventually crystallize at larger strains as discussed in the main text. 
From these stress-strain curves, stark differences in rheology are immediately present across Poisson's ratios $\nu$.
After yielding, the average stress ratio drops from an already low value of $\sim 0.17$ for $\nu = 0.4$ to a remarkably small value of $0.1$ for $\nu = -0.5$. 
Additionally, fluctuations become sharper and more rapid with decreasing $\nu$, more closely resembling avalanches or stick slip motion compared to curves at $\nu = 0.4$.

\begin{figure}
	\includegraphics[width=0.9\columnwidth]{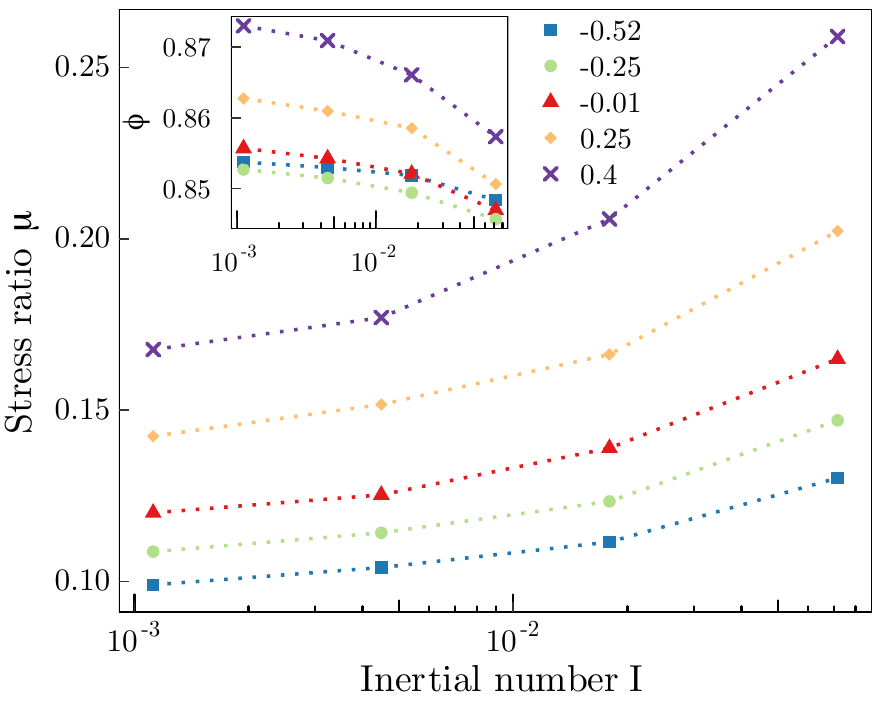}
	\caption{Stress ratio $\mu$ as a function of inertial number $I$ averaged across realizations and strains between 1.0 and 5.0 at the indicated Poisson's ratios $\nu$ and a pressure of $\tilde{P} = 0.1$. Inset plots the packing fraction $\phi$ vs. $I$.}
	\label{fig:rheology}
\end{figure}

To quantify rate effects and test for convergence to the quasistatic limit, additional strain rates of $\dot{\epsilon} \tau = 1.024 \times 10^{-3}$, $2.56 \times 10^{-4}$, and $6.4 \times 10^{-5}$ were run in addition to the rate used in the main text, $1.6 \times 10^{-5}$.
Across rates, $\mu$ is averaged in the small strain limit ($1 \le \epsilon \le 5$).
In the rigid limit, granular rheology is typically cast into a $\mu(I)$ formulation where $I$ is a dimensionless inertial number defined as $\dot{\epsilon} \langle d \rangle \sqrt{\rho/P} $ where $\langle d \rangle$ is the average grain diameter and $\rho$ is the grain density \cite{Jop2006}. 
To aid comparison to studies of traditional granular rheology, strain rates are rescaled to $I$ (Fig.~\ref{fig:rheology}).
Qualitatively, these results resemble expectations as $\mu$ drops with decreasing $I$ and gradually approaches a limit as $I$ goes to zero. 
At the strain rate presented in the main text, $\dot{\epsilon} = 1.6 \times 10^{-5} \tau^{-1}$, curves approach an asymptote suggesting the simulations are near the quasistatic limit.
Slower strain rates were not run due to excessive computational costs. 

As seen in the stress-strain curves in Fig.~\ref{fig:stress_strain}, $\mu$ decreases with decreasing $\nu$ across all rates.
Deviations across $\nu$ are exacerbated at larger $I$.
Across $\nu$, differences in packing fraction $\phi$ are also observed of about $\pm0.02$, being generally larger in the incompressible limit (inset of Fig.~\ref{fig:rheology}).
Similar variations in $\phi$ across $\nu$ were seen in unsheared samples \cite{Clemmer2024}.
At larger strains, we do see some dependence of crystallization on strain rate and note in particular that one instance of the most auxetic system ($\nu = -0.5$) crystallizing was observed at a strain rate of $6.4 \times 10^{-5}$ within a strain of 102.4 units. 

\end{document}